\documentstyle[12pt]{article}
\begin{document}

\setlength{\textheight}{240mm}
\voffset=-15mm
\baselineskip=20pt plus 2pt
\renewcommand{\arraystretch}{1.6}

\begin{center}

{\large \bf  The Quasi-localized Einstein and M{\o}ller Energy Complex as Thermodynamic Potentials}\\ 
\vspace{10mm}
I-Ching Yang \footnote{E-mail:icyang@nttu.edu.tw}

Systematic and Theoretical Science Research Group\\
and Department of Applied Science, \\
National Taitung University, Taitung 95002, Taiwan (R. O. C.) \\

\end{center}
\vspace{5mm}

\begin{center}
{\bf ABSTRACT}
\end{center}
To begin with, in this article, I obtain the Einstein and M{\o}ller energy complex in PG 
coordinates.  According to the difference of energy within region ${\mathcal M}$ between Einstein 
and M{\o}ller prescription, I could present the difference of energy of RN black hole like 
the fomula of Legendre transformation and propose that the M{\o}ller and Einstein energy 
complex play the role of internal energy and Helmholtz energy in thermodynamics.

\vspace{2mm}

\vspace{5mm}
\noindent

\newpage

\section{Introduction}

In the theory of general relativity (GR), one of the most important issues which is still 
unsolved is the localization of energy.  According to Noether's theorem, one would define 
a conserved and localized energy as a consequence of energy-momentum tensor $T^{\mu \nu}$ 
satisfying the differential conservation law 
\begin{equation}
\partial_{\nu} T^{\mu \nu} =0  .  
\end{equation}
However, in a curved space-time where the gravitational field is presented, the 
differential conservation law becomes
\begin{equation}
\nabla_{\nu} T^{\mu \nu} = \frac{1}{\sqrt{-g}} \frac{\partial}{\partial x^{\nu}}
\left( \sqrt{-g} T^{\mu \nu} \right) -\frac{1}{2} g^{\nu \rho} \frac{\partial 
g^{\nu \rho}}{\partial x^{\lambda}} T^{\mu \lambda} =0 ,
\end{equation}
and generally does not lead to any conserved quantity.  In GR, we shall look for a new 
quantity $\Theta^{\mu \nu} = \sqrt{-g} \left( T^{\mu \nu} +t^{\mu \nu} \right)$ instead 
of $T^{\mu \nu}$, which satisfies the differential conservation equation 
\begin{equation}
\partial_{\nu} \Theta^{\mu \nu} =0  , 
\end{equation}
if we want to maintain the localization characteristics of energy.  
Here, $\Theta^{\mu \nu}$ is an energy-momentum complex of matter plus gravitational 
fields and $t^{\mu \nu}$ is regarded as the contribution of energy-momentum from the 
gravitational field.  It should be noted that $\Theta^{\mu \nu}$ can be expressed as the 
divergence of the ``superpotential" $U^{\mu [\nu \rho]}$ that is antisymmetric in $\nu$
and $\rho$ as
\begin{equation}
\Theta^{\mu \nu} = {U^{\mu [\nu \rho]}}_{, \rho} . 
\end{equation}
Mathematically, it is freedom on the choice of superpotential, because one can add some 
terms $\psi^{\mu \nu \rho}$, whose divergence or double divergence is zero, to 
$U^{\mu \nu \rho}$.  A large number of definitions for the gravitational energy in GR 
have been given by many different authors, for example Einstein~\cite{1}, 
M{\o}ller~\cite{2}, Landau and Lifshitz~\cite{3}, Bergmann and Thomson~\cite{4}, 
Tolman~\cite{5}, Weinberg~\cite{6}, Papapetrou~\cite{7}, Komar~\cite{8}, Penrose~\cite{9} 
and Qadir and Sharif~\cite{10}.  On the other hand, Chang, Nester and Chen~\cite{11} 
showed that every energy-momentum complex is associated with a legitimate Hamiltonian 
boundary term and actually quasilocal.

One of those problems for using several kinds of energy-momentum complexes is that they 
may give different results for the same space-time.  Especially Virbhadra and his 
colleagues~\cite{12} showed that Einstein, Landau-Lifshitz, Papapetrou, and Weinberg 
prescriptions (ELLPW) lead to the same results in Kerr-Schild Cartesian coordinates for a 
specific class of spacetime, i.e. the general nonstatic spherically symmetric space-time 
of the Kerr-Schild class 
\begin{equation}
ds^2 = B(u,r) du^2 - 2du dr - r^2 d\Omega
\end{equation}
and the most general nonstatic spherically symmetric space-time
\begin{equation}
ds^2 = B(t,r) dt^2 - A(t,r) dr^2 -2 F(t,r) dt dr - D(t,r) r^2 d\Omega ,
\end{equation}
but not in Schwarzschild Cartesian coordinates.  Afterward Xulu~\cite{13} presented 
Bergmann-Thomson complex also ``coincides" with ELLPW complexes for a more general than 
the Kerr-Schild class metric.  Mirshekari and Abbassi~\cite{14} find a unique form for a 
special general spherically symmetric metric in which the energy of Einstein and 
M{\o}ller prescriptions lead to the same result.  In particular, whatever coordinates do 
not exist the same energy complexes associated with using definitions of Einstein and 
M{\o}ller in some space-time solutions, i.e. Reissner-Nordstr\"{o}m (RN) black hole.  On 
the other hand, Yang and Radinschi~\cite{15} attemptd to investigate the difference 
between the energy of Einstein prescription $E_{\rm Einstein}$ and M{\o}ller prescription 
$E_{\rm M{\o}ller}$, and observed the difference $\Delta E =E_{\rm Einstein} - 
E_{\rm M{\o}ller}$ can be related to the energy density of the matter fields $T^0_0$ as
\begin{equation}
\Delta E \sim  r^3 \times T^0_0 .
\end{equation}
Matyjasek~\cite{16} also presented two analogous relations which are 
\begin{equation}
\Delta E = 4\pi r^3 T^0_0 
\end{equation} 
for the simplified stress-energy tensor of the matter field and
\begin{equation}
\Delta E = 4\pi r^3 {\langle T^r_r \rangle}^{(s)}_{ren}
\end{equation} 
for the approximate renormalized stress-energy tensor of the quantized massive scalar 
$(s=0)$, spinor $(s=1/2)$ and vector $(s=1)$ field.  Later, Vagenas~\cite{17} hypothesized 
that $\alpha^{(\rm Einstein)}_n$ and $\alpha^{(\rm M{\o}ller)}_n$ are the expansion 
coefficients of $E_{\rm Einstein}$ and $E_{\rm M{\o}ller}$ in the inverse powers of $r$, 
and found out an interesting relation between these two coefficients
\begin{equation}
\alpha^{(\rm Einstein)}_n = \frac{1}{n+1} \alpha^{(\rm M{\o}ller)}_n  .
\end{equation}
Finally, Matyjasek~\cite{16} and Yang {\it et. al.}~\cite{18} pointed out the following 
formula respectively
\begin{equation}
E_{\rm M{\o}ller} = E_{\rm Einstein} - r \frac{d E_{\rm Einstein}}{dr} .
\end{equation}
It should be noted that these relations in Eq. (7)-(11) offer us the mathematical
formula between $E_{\rm Einstein}$ and $E_{\rm M{\o}ller}$ only.  The remainder of the 
article is organized as follows.  In section 2, I will calculate the energy distribution 
for generalized Painlev\'{e}-Gullstrand (PG) coordinates~\cite{19} by using the Einstein 
and M{\o}ller complex.  In section 3, the physical explanation of the difference 
$\Delta E$ will be given.  I will summarize and conclude finally in section 4.  In this 
article, I use geometrized units in which $c=G=\hbar =1$ and the metric has signature 
$(+---)$.

\section{Using the Einstein and M{\o}ller Energy Complex in generalized PG coordinates}

The continuation of black holes across the horizon is a well understood problem discussed 
on GR.  The difficulties of the Schwarzschild coordinates $(t, r, \theta, \phi)$ at the 
horizons of a nonrotating black hole provide a vivid illustration of the fact that the 
meaning of the coordinates is not independent of the metric tensor $g_{\mu \nu}$ in GR.  
Several coordinate systems produce a metric that is manifestly regular at horizons, i.e. 
the Kruskal-Szekeres, Eddington-Finkelstein, and PG coordinates.  However, PG coordinates 
have often been employed to study the physics of black holes.  They have been applied to 
analyse quantum dynamical black holes~\cite{20}, and used extensively in derivations of 
Hawking radiation as tunneling following the work of Parikh and Wilczek~\cite{21}.  In  
this section, while using PG coordinates, I will find out the energy of static spherically 
symmetric black hole solutions in Einstein and M{\o}ller prescriptions.  In 
four-dimensional theory of gravity, I can write the static spherically symmetric metrics 
in the form  
\begin{equation}
ds^2 = f dt^2 -f^{-1} dr^2 -r^2 d\Omega ,
\end{equation}
where $f$ is a function of $r$, i.e. $f=f(r)$.  Let me transform to generalized PG 
coordinates~\cite{19} and introduce the PG time $ dt_p = dt + 
\beta dr$, thus 4-metric can be written as
\begin{equation}
ds^2 = f dt_p^2 - 2\sqrt{1 -\frac{f}{A^2}} dt_p dr -\frac{1}{A^2} dr^2 -r^2 d\Omega ,
\end{equation}
where $A \equiv \sqrt{f /(1 -f^2 \beta^2)}$.

At the outset, the energy component in the Einstein prescription~\cite{1} is given by
\begin{equation}
E_{\rm Einstein} = \frac{1}{16\pi} \int \frac{\partial H^{0l}{0}}{\partial x^l} d^3 x ,
\end{equation}
where $H^{0l}_0$ is the corresponding von Freud superpotential
\begin{equation}
H^{0l}_0 = \frac{g_{0n}}{\sqrt{-g}} \frac{\partial}{\partial x^m} \left[ (-g) (g^{0n} g^{lm}
- g^{ln} g^{0m} ) \right] ,
\end{equation}
and the Latin indices take values from 1 to 3.  For performing the calculations concering the 
energy component of the Einstein energy-momentum complex, I have to transform the 
spatial parts of above metric (13) into the quasi-Cartesian coordinates $(x, y, z)$
\begin{eqnarray}
ds^2 & = & A^2 dt^2_p - 2 \sqrt{1 -\frac{f}{A^2}} dt_p (\frac{x}{r} dx + \frac{y}{r} dy 
+ \frac{z}{r} dz )  \nonumber \\
 &  & - (\frac{1}{A^2} -1) (\frac{x}{r} dx + \frac{y}{r} dy + \frac{z}{r} dz )^2
- (dx^2 + dy^2 + dz^2) .
\end{eqnarray}
Then, the required nonvanishing components of the Einstein energy-momentum complex
$H^{0l}_0$ are 
\begin{eqnarray}
H^{01}_0 & = & \frac{2Cx}{r} , \nonumber \\
H^{02}_0 & = & \frac{2Cy}{r} , \nonumber \\
H^{03}_0 & = & \frac{2Cz}{r} , \nonumber \\
\end{eqnarray}
and these are easily shown in spherical coordinates to be a vector
\begin{equation}
H^{0r}_0 = \frac{2C}{r}  \hat{r} , 
\end{equation}
where $C= 1-f$ and $\hat{r}$ is the outward normal. Applying the Gauss therorem I obtain
\begin{equation}
E_{\rm Einstein} = \frac{1}{16 \pi} \oint H^{0r}_0 \cdot \hat{r} r^2 d \Omega ,
\end{equation}
and the integral being taken over a sphere of radius $r$ and the differential solid angle 
$d \Omega$.  The Einstein energy complex within radius $r$ reads
\begin{equation}
E_{\rm Einstein} = \frac{r}{2} ( 1-f) .
\end{equation}

Next, the energy component of the M{\o}ller energy-momentum complex~\cite{2} is described
as
\begin{equation}
E_{\rm M{\o}ller} = \frac{1}{8\pi} \int \frac{\partial \chi_0^{0l}}
{\partial x^{l}} d^3 x ,
\end{equation}
where $\chi_0^{0l}$ is the M{\o}ller superpotential
\begin{equation}
\chi_0^{0l} = \sqrt{-g} \left( \frac{\partial g_{0 \alpha}}{\partial x^{\beta}} - 
\frac{\partial g_{0 \beta}}{\partial x^{\alpha}} \right) g^{0 \beta} g^{l \alpha},
\end{equation}
and the Greek indices run from 0 to 3.  However, the only nonvanishing component of 
M{\o}ller's superpotential is
\begin{equation}
\chi_0^{01} = \frac{df}{dr} r^2 \sin \theta .
\end{equation}
Applying the Gauss theorem, I evaluate the integral over the surface of a sphere within
radius $r$, and find the energy distribution is 
\begin{equation}
E_{\rm M{\o}ller} = \frac{r^2}{2} \frac{df}{dr} .
\end{equation}

Here, I consider the results of calculation for two cases of the simplest black hole 
solutions, i.e. Schwarzschild and RN solution.  In the first case I have $f=1 -2M/r$, 
therefore the energy complex of Einstein is 
\begin{equation}
E_{\rm Einstein} = M ,
\end{equation}
and of M{\o}ller is also
\begin{equation}
E_{\rm M{\o}ller} = M .
\end{equation}
For the next case it is defined that $f=1 -2M/r +Q^2/r^2$, so the energy complex in
Einstein prescription is 
\begin{equation}
E_{\rm Einstein} = M -\frac{Q^2}{2r} ,
\end{equation}
and in M{\o}ller prescription is 
\begin{equation}
E_{\rm Einstein} = M -\frac{Q^2}{r} .
\end{equation}
It should be noted that the above results of Einstein energy complex in PG Cartesian
coordinates are equivalent to in Schwarzschild Cartesian and Kerr-Schild Cartesian 
ones~\cite{12}, but the time coordinate of these three coordinates is different to each 
other.  Using the Schwarzschild black hole as an example, the time coordinate of 
Schwarzschild Cartesian coordinates $t$, of Kerr-Schild Cartesian coordinates 
\begin{equation}
v = t + r + 2M \ln \left| \frac{r}{2M} -1 \right|  ,
\end{equation}
and of PG Cartesian coordinates 
\begin{equation}
t_p = t + 4M \left( \sqrt{\frac{r}{2M}} + \frac{1}{2} \ln \left| \frac{\sqrt{r/2M} -1}
{\sqrt{r/2M} +1} \right| \right) 
\end{equation}
are not the same.  In other words, independent of the choices of these three kinds of 
time coordinate, the energy complex of Einstein within radius $r$ is $E_{\rm Einstein}
=M$.  It is indefinite that  the energy complex of Einstein is universal for any kinds of 
time coordinate. Some quasi-local energy expressions~\cite{22} and the Einstein 
energy-momentum pseudotensors are coordinate-independent in spherically symmetric 
space-time.  It remains to investigate whether the coordinate-independent is the property of
the spherically symmetric space-time.

\section{Legendre transformation between the Einstein's and M{\o}ller's Energy Complex}

To understand the physical meaning of difference $\Delta E$, let me to begin with examining 
the RN black hole, which is a static spherically symmetric solution with two horizons, as an 
example. The line element of RN black hole can be written as
\begin{equation}
ds^2 = f(r) dt^2 - f^{-1} (r) dr^2 -r^2 d\Omega ,
\end{equation}
where
\begin{equation}
f(r) = \left( 1- \frac{r_+}{r} \right) \left( 1- \frac{r_-}{r} \right) ,
\end{equation}
$r_+ = M +\sqrt{M^2-Q^2}$ is the event horizon and $r_-  = M - \sqrt{M^2-Q^2}$ is the 
inner Cauchy horizon.  According to Eq.(19) and Eq.(23), the Einstein energy complex with 
radius $r$ is
\begin{equation}
E_{\rm Einstein} = \frac{r_+ + r_-}{2} - \frac{r_+ r_-}{2r} ,
\end{equation}
and the M{\o}ller energy complex is 
\begin{equation}
E_{\rm M{\o}ller} = \frac{r_+ + r_-}{2} - \frac{r_+ r_-}{r} .
\end{equation}
Therefore, the difference of energies with radius $r$ between the Einstein and M{\o}ller 
prescription can be obtained as
\begin{equation}
\Delta E  = \frac{r_+ r_-}{2r}  .
\end{equation}
In the article of Nester {\it et. al.}~\cite{11}, they had stated that {\it ``Consequently, 
there are various of energy, each corresponding to a different choice of boundary condition; 
this situation can be compared with thermodynamics with its various energies: internal, 
enthalpy, Gibbs, and Helmholtz."}  Hence, I insert the idea of black hole thermodynamics 
to compare energy-momentum complex with thermodynamic potential.

Afterward, I would introduce two thermodynamic qualities of black hole, the Hawking 
temperature~\cite{19} 
\begin{equation}
T_H = \frac{1}{4\pi } \left. \frac{\partial f}{\partial r} \right|_{r_h}  
\end{equation}
and the Bekenstein-Hawking entropy~\cite{23} 
\begin{equation}
S_{BH} = \left. \frac{\cal {A}}{4} \right|_{r_h} = \pi r_h^2 .
\end{equation}
Because those two qualities are only defined on event horizon, at $r = r_+$, the 
temperature is given as
\begin{equation}
T^{+} = \frac{r_+ - r_-}{4\pi r_+^2}  = \frac{\sqrt{M^2-Q^2}}{2\pi (M+ \sqrt{M^2-Q^2})^2}
\end{equation}
and the entropy is also given as
\begin{equation}
S^{+} = \pi r_+^2  = \pi (M+ \sqrt{M^2-Q^2})^2,
\end{equation}
Supposing that we consider the region between those two horizons, shown as 
${\mathcal{M}} = {\cal{B}}^3 (r_+) - {\cal{B}}^3 (r_-)$, the difference of energies will 
be obtained in the form
\begin{equation}
{\Delta E \left| \right.}^{r=r_+}_{r=r_-} = -\frac{r_+ - r_-}{2} = -\sqrt{M^2-Q^2} = -2 T^+ S^+ .
\end{equation}
Here, ${\cal{B}}^3 (r)$ is a 3-sphere within a radius $r$.  The term $T^+ S^+ $ can be considered
that the heat flow streams out the region ${\mathcal{M}}$ by passing the boundary of 
${\cal{B}}^3 (r_+)$.  Therefore, Eq. (39) would be rewitten as
\begin{equation}
\left. E_{\rm M{\o}ller} \right|^{r_+}_{r_-} -\left. E_{\rm Einstein} \right|^{r_+}_{r_-} 
= 2 T^+ S^+  .
\end{equation}
It is meaning that the difference of energies between Einstein and M{\o}ller prescription equal to the  
double of the heat flow stearms out by passing the bounday of ${\cal{B}}^3 (r_+)$ in the region 
${\mathcal{M}}$ of RN black hole.

On the other hand,  to base on Zhao's study~\cite{24},  the entropy of black hole, which has two
horizons, is defined as $\tilde{S}= S^+ + S^-$, where the entropy of the inner Cauchy horizon can
be shown as 
\begin{equation}
S^- = \pi r^2_- = \pi \left( M - \sqrt{M^2 - Q^2} \right) ^2 ,
\end{equation}
and the temperature of the inner Cauchy horizon is given as 
\begin{equation}
T^- = \frac{\kappa_-}{2\pi} ,
\end{equation}
where the surface gravity of the inner Cauchy horizon is~\cite{25} 
\begin{equation}
\kappa_- = \lim_{r \rightarrow r_-} - \frac{1}{2(r - r_-)}  \sqrt{-\frac{g^{11}}{g^{00}}}
= \frac{r_+ - r_-}{2 r_-^2} .
\end{equation}
So the difference of energy between the Einstein prescription and M{\o}ller prescription within 
the region ${\mathcal{M}}$ can be written as
\begin{equation}
\left. E_{\rm M{\o}ller} \right|^{r_+}_{r_-} -\left. E_{\rm Einstein} \right|^{r_+}_{r_-} 
= T^+ S^+ + T^- S^-  ,
\end{equation}
and the heat flow will be with respect to both two boundaries of ${\mathcal{M}}$.  To rewrite 
Eq. (44) as
\begin{equation}
\left. E_{\rm Einstein} \right|_{\mathcal{M}} = \left. E_{\rm M{\o}ller} \right|_{\mathcal{M}} 
- \sum_{\partial {\mathcal{M}}} T S  ,
\end{equation}
these heat flows are exhibited on every boundary of ${\mathcal{M}}$.  Comparing Eq. (45) 
with the Legendre transformation,  $E_{\rm M{\o}ller}$ and $E_{\rm Einstein}$ in the region
${\mathcal{M}}$ play the role of internal energy $U$ and Helmholtz energy $F$ in 
thermodynamics, even so there is a puzzle where $E_{\rm M{\o}ller}$ or $E_{\rm Einstein}$ 
are not a function of $T$ or $S$.  We could obtain not only a physical meaning of the 
difference of energies in Eq. (34),  although the statement can only be used to RN black hole, 
but also such a result agreed with the entropy redefined in Zhao's article.

\section{Conclusion and Discussion}
I have attempted to answer two questions in this article.  One is whether the calculation
of Einstein energy-momentum complex is acceptable in PG coordinate, and the other is whether 
those energy-momentum complex can be described as a thermodynamic potential.  Here, the 
expression for energy of the static spherically symmetric space-time with the PG Cartesian
coordinates, Eq.(19), is obtained $ E_{\rm Einstein} = (1-f)r/2 $. This is a reasonable and 
satisfactory result, because Virbhadra~\cite{12}, using the Kerr-Schild Cartesian coordinates, 
and Yang {\it et al.}~\cite{18}, using the Schwarzschild Cartesian coordinates, also got the same 
expression.  It is interesting to investigate whether there is any coincidence between the energy 
expressions with those three time coordinate.  

In addition, I have showed that the relational formula about $E_{\rm Einstein}$ and 
$E_{\rm M{\o}ller}$ is similar to the Legendre transformation.  The $E_{\rm M{\o}ller}$ 
and $E_{\rm Einstein}$ are regarded as the equivalent of internal energy $U$ and 
Helmholtz energy $F$ in the region $\mathcal{M}$. Although, the transformation takes
us from a function of one pair variables to the other.  It means that $S_{BH}$ and $T_H$
must be the variable of $E_{\rm M{\o}ller}$ and $E_{\rm Einstein}$, but I do not verify 
that yet.  On the other hand, when I set ${\mathcal{S}}_{\perp} = \pi r^2$ to be a 
variable, the second term in the right-hand side of Eq.(11) can be replaced as 
\begin{equation}
E_{\rm M{\o}ller} = E_{\rm Einstein} - 2 {\mathcal{S}}_{\perp} \frac{d E_{\rm Einstein}}
{d {\mathcal{S}}_{\perp}} .
\end{equation}
To compare with $F=U-TS$, we could obtain 
\begin{equation}
T_H = \left. \frac{d E_{\rm Einstein}}{d {\mathcal{S}}_{\perp}} \right|_{r_h} .
\end{equation}
Here the formula of Eq.(44) presents that $E_{\rm Einstein}$ and $E_{\rm M{\o}ller}$ play 
the role of $U$ and $F$, and is opposite to the view of above.   
  
In summary, I have obtained the Einstein and M{\o}ller energy complexes of static 
spherically symmetric black hole with generalized PG coordinates in which has been used
in derivations of Hawking radiation as tunneling.  Base on the calculation of energy 
expression in generalized PG coordinates, in Eq.(45) I have combined the difference 
$\Delta E$ with the temperature and entropy of black hole, but Eq.(4g) do not fit  
in with the Legendre transformation. Nevertheless, it is an example to show that the 
energy-momentum complexes of RN black hole will compare with thermodynamic 
potential, and future research should be considered on more kinds of space-time. 

\vspace{10mm}
{\bf ACKNOWLEDGMENTS} \\
I would like to thank Prof. Chopin Soo and Prof. Su-Long Nyeo for useful suggestions
and discussions.  This work is partially supported by National Center for Theoretical
Sciences, Taiwan.

\end{document}